\documentclass[onecolumn,showpacs,floatfix,10pt,nofootinbib]{revtex4}
\usepackage{amssymb,amsmath}
\usepackage{graphicx,float}
\usepackage{indentfirst}
\newcommand{\be}{\begin{equation}}
\newcommand{\ee}{\end{equation}}

\newcommand{\ba}{\begin{eqnarray}}
\newcommand{\ea}{\end{eqnarray}}

\newcommand{\lie}{\pounds_{\bf n}}
\newcommand{\el}{^}
\begin{document}

\title{Black string corrections in variable tension braneworld scenarios}

\author{Rold\~ao da Rocha}
\email{roldao.rocha@ufabc.edu.br}
\affiliation{Centro de Matem\'atica, Computa\c c\~ao e Cogni\c c\~ao, Universidade Federal do ABC (UFABC) 09210-170, Santo Andr\'e, SP, Brazil.}
\author{J. M. Hoff da Silva}
\email{hoff@feg.unesp.br;hoff@ift.unesp.br}
\affiliation{Departamento de F\1sica e Qu\1mica, Universidade
Estadual Paulista, Av. Dr. Ariberto Pereira da Cunha, 333,
Guaratinguet\'a, SP, Brazil.}


\begin{abstract}Braneworld models with variable tension are investigated, and the corrections on the black string horizon along the extra dimension are provided. Such corrections are encrypted in additional terms involving the covariant derivatives of the variable tension on the brane, providing profound consequences concerning the black string horizon variation along the extra dimension, near the brane. The black string horizon behavior is shown to be drastically modified by the terms corrected by the brane variable tension. In particular, a model motivated by the phenomenological interesting case regarding E\"otv\"os branes is investigated. It forthwith provides further physical features regarding variable tension braneworld scenarios, heretofore concealed in all previous analysis in the literature. All precedent analysis considered uniquely the expansion of the metric up to the second order along the extra dimension, what is able to evince solely the brane variable tension absolute value. Notwithstanding, the expansion terms aftermath, further accomplished in this paper from the third order on, elicits the successive covariant derivatives of the brane variable tension, and their respective coupling with the extrinsic curvature, the Weyl tensor, and the Riemann and Ricci tensors, as well as the scalar curvature. Such additional terms are shown to provide sudden modifications in the black string horizon in a variable tension braneworld scenario.\end{abstract}\maketitle
\flushbottom


\section{Introduction}

Braneworld models perform a distinct branch of high energy physics. Several of the main ideas concerning such models were inspired in formal advances in string theory \cite{HW}. It is possible to delineate the precursory works dealing with the possibility of a braneworld universe \cite{PREV}, and nowadays there is a great variety of alternative models dealing with several aspects regarding braneworlds. Motivated by a comprehensive program regarding gravity and black strings on braneworld scenarios, and subsequent generalizations probing extra-dimensional features \cite{nossos,adsbranes,prd2010,ROLDAO/CARLAO,nova,Gergely:2006hd,loo,Anderson:2005af}, some gravitational aspects concerning black strings, arising from braneworld models with variable brane tension, are introduced and widely investigated. In order to analyze  a conceivable gravitational effect, arising genuinely from the brane variable tension, we analyze the corrections on the black string time-dependent horizon, in such scenario. This formalism provides manageable models and their possible ramifications into some aspects of gravity in this context, cognizable corrections, and physical effects as well. Besides the black string behavior is sharply modified by the variable tension brane localized at $y=0$ (the extra dimension is denoted by $y$),  unexpected additional terms in a Taylor expansion of the metric along the extra dimension are elicited. This Taylor expansion regards the black string horizon behavior  along the extra dimension. Taking into account a variable tension brane, this expansion involves the covariant derivatives of the brane tension -- besides its Lie derivative along the extra dimension likewise. Such expansion for instance was used in \cite{casadio1} to investigate the  gravitational collapse of compact objects in braneworld scenarios.

Braneworld models, wherein a brane is endowed with variable tension $\lambda = \lambda(x^\mu)$, are a prominent approach as an attempt to ascertain new signatures coming from high-energy physics \cite{VACARU}. In fact, due to the drastic modification of the temperature of the universe along its cosmological evolution, a variable tension braneworld scenario is indeed demanded. The full covariant variable tension brane dynamics was established in Ref. \cite{GERGELY2008}, and further explored in \cite{GERGELY2009}. Moreover,  the variable tension was implemented in the braneworld model consisting of two branes \cite{PRD,PRDII}, also in the context of scalar tensor bulk gravity \cite{JMHEP}. Furthermore, the cosmological evolution of the universe was investigated in a particular model which the brane tension has an exponential dependence with the scale factor \cite{EPJC}.

In this paper we are concerned to analyze the information about the black string behavior, evinced by a time variable tension on the brane, delving into the Taylor expansion outside a black hole metric along the extra dimension, where the corrections in the area of the 5-dimensional black string horizon are elicited.  It is shown how the variable tension and its covariant derivatives determine the variation in the area of the black string horizon along the extra dimension. It induces interesting physical effects, for instance the preclusion of the horizon decreasing, when the brane tension varies in a model physically motivated by E\"otv\"os law.

Our program throughout this paper explicitly consists of the following: the next Section provides the black string behavior along the extra dimension, studying the variation in the black string horizon due to terms including the covariant derivatives of the variable brane tension. In particular, it is considered the E\"otv\"os phenomenological brane case, and the Lie derivatives of $\lambda$ along the extra dimension are identically zero and therefore concealed in the subsequent analysis. Some results \cite{maartens} are extended in order to encompass this case. The fine character of the expansion along the extra dimension is crucial to analyze variable tension models. The Taylor expansion along the extra dimension, concerning variable tension braneworld models,  provides  terms of superior order and  brings some more precise information about the behavior of the black string than in models with brane constant tension. Furthermore, when variable tension braneworld models are taken into account, the extra terms ---  from the third order on --- probe physical features concerning the variable tension. For instance, such terms induce the horizon to decrease in a different rate along the extra dimension, when we analyze the physical concrete example of E\"otv\"os branes. In Section III we delve into the corrections for the Schwarzschild case analysis, and to this point the framework is completely general. In Section IV the specific and physically motivated example, regarding E\"otv\"os branes endowed with a de Sitter-like scale factor variable tension, is investigated.

\section{black string behavior along the extra dimension}
In this Section the comprehensive formalism in \cite{maartens} is briefly introduced and reviewed.
Hereon $\{\theta_\mu\}$, {\footnotesize{$\mu = 0,1,2,3$}} [$\{\theta_A\}$, {\footnotesize{$A=0,1,2,3,4$}}] denotes a basis for the cotangent space $T^\ast_xM$ at a point $x$ in a 3-brane
$M$ embedded in a bulk.
If a local coordinate chart is chosen it is possible to represent $\theta^A = dx^A$.
Take now $n = n^Ae_A$ a time-like vector orthogonal to $T^\ast_xM$ and let $y$ be the associated Gaussian coordinate, indicating
how an observer upheavals out the brane into the bulk. In particular, $n_Adx^A = dy$ in the hyper-surface defined by $y=0$.
A vector field $v = x^Ae_A$ in the  bulk  is split into components in the brane and orthogonal to the brane, respectively as
 $v = x^\mu e_\mu + ye_4 = (x^{\mu},y)$. The bulk is endowed with a
metric $\mathring{g}_{AB}dx^A dx^B = g_{\mu\nu}(x^\alpha,y)\,dx^\mu dx^\nu + dy^2$. The brane metric components $g_{\mu\nu}$ and the bulk metric are related by
$
\mathring{g}_{\mu\nu} = g_{\mu\nu} + n_\mu n_\nu,
$ since according the notation in \cite{maartens}, as $g_{44} = 1$ and $g_{i4} = 0$, the bulk indices $A,B$ effectively run from 0 to 3.

It is well know that the 4-dimensional gravitational constant is
an effective coupling constant inherited from the fundamental
coupling constant, and the 4-dimensional cosmological constant is
non zero when the  balance between the bulk cosmological constant
and the brane tension --- provided by the Randall-Sundrum
braneworld model \cite{RSI} --- is broken \cite{maartens}:
\be\kappa^2_{4}=\frac{1}{6}\lambda\kappa^4_5,\qquad\qquad\quad
\Lambda_4=\frac{\kappa_5^{2}}{2}\Big(\Lambda_{5}+\frac{1}{6}\kappa_5^{2}\lambda^{2}\Big),\ee
where $\Lambda_4$ is the effective brane cosmological constant,
$\kappa_5$ [$\kappa_4$] denotes the 5-dimensional [4-dimensional]
gravitational coupling, and $\lambda$ is the brane tension. The
extrinsic curvature $K_{\mu\nu} = \frac{1}{2}\lie g_{\mu\nu}$
(hereon $\lie$ denotes  the Lie derivative, which in Gaussian
normal coordinates reads $ {\bf \pounds}_{\bf n}=\partial/\partial
y$). The junction condition determines the extrinsic curvature on
the brane, as
 \be\label{ext}
K_{\mu\nu}=-\frac{1}{2}\kappa_5^2 \left[T_{\mu\nu}+ \frac{1}{3}
\left(\lambda-T\right)g_{\mu\nu} \right] \,,
 \ee
where $T=T^\mu{}_\mu $. In what follows we denote $K = K_\mu^{\;\,\mu}$ and $K^2 = K_{\alpha\beta}K^{\alpha\beta}$.

Given the 5-dimensional Weyl tensor
\begin{equation}
 C_{\mu\nu\sigma\rho} = R_{\mu\nu\sigma\rho} - \frac{2}{3} (\mathring{g}_{[\mu\sigma} R_{\nu]\rho} + \mathring{g}_{[\nu\rho} R_{\mu]\sigma}) - \frac{1}{6} R ( \mathring{g}_{\mu[\sigma} \mathring{g}_{\nu\rho]}),
\end{equation}\noindent where $R_{\mu\nu\sigma\rho}$ denotes the components of the bulk Riemann tensor ($R_{\mu\nu}$ and $R$ obviously are the associated Ricci tensor and the scalar curvature), the symmetric and trace-free components respectively denoted by ${\cal E}_{\mu\nu}$  and ${\cal B}_{\mu\nu\alpha}$  are known as
 the electric and magnetic Weyl tensor components --- given by ${\cal E}_{\mu\nu} = C_{\mu\nu\sigma\rho} n^\sigma n^\rho$ and ${\cal B}_{\mu\nu\alpha} = g_\mu^{\;\rho} g_\nu^{\;\sigma}
C_{\rho\sigma\alpha\beta}n^\beta$.
 The Weyl tensor represents the part of the curvature that is not determined locally by matter, and the field equations for the Weyl tensor are simulated by Bianchi identities, determining the part of the spacetime curvature that depends on the matter distribution at other points
\cite{1}.

The effective field equations are complemented by the a set of equations, that are forthwith obtained from the 5-dimensional Einstein and Bianchi equations \cite{GCGR}. Those equations are also obtained in \cite{Gergely:2003pn,maartens}, as well as in a brane with variable tension context \cite{GERGELY2008}. All approaches are shown to be completely similar and hereon the following effective field equations are considered:
\begin{eqnarray}
 {\bf \pounds}_{\bf n} K_{\mu\nu}&=& K_{\mu\alpha}K^\alpha
{}_\nu - {\cal E}_{\mu\nu}-\frac{1}{6}\Lambda_5 g_{\mu\nu} \label{exp1}\\
{\bf \pounds}_{\bf n} {\cal E}_{\mu\nu}  &=& \nabla^\alpha
{\cal B}_{\alpha(\mu\nu)} + \frac{1}{6}
\Lambda_5\left(K_{\mu\nu}-g_{\mu\nu}K\right)
+K^{\alpha\beta}R_{\mu\alpha\nu\beta} +3K^\alpha{}_{(\mu}{\cal
E}_{\nu)\alpha}-K{\cal E}_{\mu\nu}\nonumber\\&&
\qquad\qquad+\left(K_{\mu\alpha}K_{\nu\beta}
-K_{\alpha\beta}K_{\mu\nu}\right)K^{\alpha\beta} \label{exp2}
\\  {\bf \pounds}_{\bf n} {\cal B}_{\mu\nu\alpha}&=&-
2\nabla_{[\mu}{\cal E}_{\nu]\alpha}+K_\alpha{}^\beta {\cal
B}_{\mu\nu\beta} -2{\cal B}_{\alpha\beta [\mu }K_{\nu]}{}^\beta
\label{exp3}
\\ {\bf \pounds}_{\bf n} R_{\mu\nu\alpha\beta}
&=&-2R_{\mu\nu\gamma [\alpha}K_{\beta]}{}^\gamma
-\nabla_{\mu}{\cal B}_{\alpha\beta\nu} + \nabla_{\mu}{\cal
B}_{\beta\alpha\nu}. \label{exp4}
\end{eqnarray}
These equations are to be solved subject to the boundary
condition $
 {\cal B}_{\mu\nu\alpha} = 2\nabla_{[\mu}K_{\nu]\alpha}$ at the brane \cite{maartens}.

Those expressions are subsequently used in order to explicitly calculate the terms of the Taylor expansion of the metric along the extra dimension. Such expansion provides the black string profile and some physical consequences.
The equations above were used to develop a covariant analysis of the weak field~\cite{GCGR}, and are used to develop a Taylor expansion of the metric along the extra dimension, which predicts the black string behavior. We intend here to perform the calculations beyond second order term in the expansion along the extra dimension. Solely in this way it is possible to realize how the variable tension generates additional terms concerning the covariant derivatives of the brane variable tension.

The standard Taylor expansion along the extra dimension $y$ is given by the expression
\begin{eqnarray}
\hspace*{-0.3cm}g_{\mu\nu}(x,y)&=& g_{\mu\nu}(x,0) +\left({\bf \pounds}_{\bf n}g_{\mu\nu}(x,y)\right)\vert_{y=0}\,|y| +\left({\bf \pounds}_{\bf n}\left({\bf \pounds}_{\bf n}g_{\mu\nu}(x,y)\right)\right)\vert_{y=0}\,\frac{|y|^2}{2!} \nonumber\\
&&+ \left({\bf \pounds}_{\bf n} \left({\bf \pounds}_{\bf n}\left({\bf \pounds}_{\bf n}g_{\mu\nu}(x,y)\right)\right)\right)\vert_{y=0}\,\frac{|y|^3}{3!} + \cdots + (\lie^k (g_{\mu\nu}(x,y))\vert_{y=0} \frac{|y|^k}{k!} + \cdots\label{liee}
\end{eqnarray}
 The term in the first order $|y|$ above 
is immediately calculated by the definition of the extrinsic curvature $K_{\mu\nu} = \frac{1}{2}\lie g_{\mu\nu}$ and by the junction condition (\ref{ext}).  The term in $y^2$ is proportional to $\lie K_{\mu\nu}$, which is given by Eq.(\ref{exp1}), and in order to calculate the term $K_{\mu\alpha}K^\alpha
{}_\nu$ in Eq.(\ref{exp1}), the junction condition (\ref{ext}) is used forthwith. Furthermore, the coefficient term of $|y|^3$ in Eq.(\ref{liee})
can be expressed as
\begin{equation} 2{\bf \pounds}_{\bf n}\left({\bf \pounds}_{\bf n}K_{\mu\nu}\right)\frac{|y|^3}{3!} 
=\left({\bf \pounds}_{\bf n}\left(K_{\mu\alpha}K^\alpha
{}_\nu\right) - {\bf \pounds}_{\bf n}{\cal E}_{\mu\nu}-\frac{\Lambda_5}{6}\lie g_{\mu\nu}\right)\frac{|y|^3}{3},\label{ult}\end{equation}\noindent where Eq.(\ref{exp1}) was used. As the Lie derivatives terms in the right hand side of this last expression are respectively given by Eq.(\ref{exp1}) (where the Leibniz rule ${\bf \pounds}_{\bf n}\left(K_{\mu\alpha}K^\alpha
{}_\nu\right) = {\bf \pounds}_{\bf n}\left(K_{\mu\alpha}\right)K^\alpha
{}_\nu + K_{\mu\alpha}{\bf \pounds}_{\bf n}\left(K^\alpha
{}_\nu\right)$ is employed), by Eq.(\ref{exp2}), and by the definition of the extrinsic curvature, one arrives further at the expression for $|y|^3$ in Eq.(\ref{liee}). Finally, the term in  $y^4$ is obtained when the Lie derivative of the right hand side of Eq.(\ref{ult}) is taken into account, as well as Eqs.(\ref{exp1})-(\ref{exp4}) again.

Explicitly, up to fourth order in the extra dimension, the Taylor expansion is given by (hereon we denote $g_{\mu\nu}(x,0) = g_{\mu\nu}$):
 \ba
&&
\hspace*{-0.4cm}g_{\mu\nu}(x,y)= g_{\mu\nu}(x,0)-\kappa_5^2\left[
T_{\mu\nu}+\frac{1}{3}(\lambda-T)g_{\mu\nu}\right]\,|y| \nonumber\\
&&~~{}+\left[-{\cal E}_{\mu\nu} +\frac{1}{4}\kappa_5^4\left(
T_{\mu\alpha}T^\alpha{}_\nu +\frac{2}{3} (\lambda-T)T_{\mu\nu}
\right) +\frac{1}{6}\left( \frac{1}{6}
\kappa_5^4(\lambda-T)^2-\Lambda_5
\right)g_{\mu\nu}\right]\, y^2+ \nonumber\\
&& +\left.\Bigg[2K_{\mu\beta}K^{\beta}_{\;\,\alpha}K^{\alpha}_{\;\,\nu} - ({\cal E}_{\mu\alpha}K^{\alpha}_{\;\,\nu}+K_{\mu\alpha}{\cal E}^{\alpha}_{\;\,\nu})-\frac{1}{3}\Lambda_5K_{\mu\nu}-\nabla^\alpha{\cal B}_{\alpha(\mu\nu)} + \frac{1}{6}
\Lambda_5\left(K_{\mu\nu}-g_{\mu\nu}K\right)
 \right.\nonumber\\
&&\left.+K^{\alpha\beta}R_{\mu\alpha\nu\beta}+3K^\alpha{}_{(\mu}{\cal
E}_{\nu)\alpha}-K{\cal E}_{\mu\nu}+\left(K_{\mu\alpha}K_{\nu\beta}
-K_{\alpha\beta}K_{\mu\nu}\right)K^{\alpha\beta}-\frac{\Lambda_5}{3}K_{\mu\nu}\Bigg]\;\frac{|y|^3}{3!} +\right. \nonumber\\
&&+\left.\Bigg[\frac{\Lambda_5}{6}\left(R-\frac{\Lambda_5}{3} + K^2\right)g_{\mu\nu} + \left(\frac{K^2}{3}- \Lambda_5\right)K_{\mu\alpha}K^{\alpha}_{\;\,\nu} + (R-\Lambda_5 + 2K^2){\cal E}_{\mu\nu}\right.\nonumber\\&+&\left. \left(K^{\alpha}_{\;\,\sigma}K^{\sigma\beta} + {\cal E}^{\alpha\beta} +KK^{\alpha\beta}\right)\,R_{\mu\alpha\nu\beta} - \frac{1}{6}\Lambda_5R_{\mu\nu}
 + 2 K_{\mu\beta}K^{\beta}_{\;\,\sigma}K^\sigma_{\;\,\alpha}K^\alpha_{\;\,\nu} + K_{\sigma\rho}K^{\sigma\rho}K\,K_{\mu\nu}\right.\nonumber\\&+&\left.
  {\cal E}_{\mu\alpha}\left(K_{\nu\beta}K^{\alpha\beta}-3K^\alpha_{\;\,\sigma}K^{\sigma}_{\;\,\nu} + \frac{1}{2}KK^\alpha_{\;\,\nu}\right)+\left(\frac{7}{2}KK^\alpha_{\;\,\mu}- 3K^\alpha_{\;\,\sigma}K^\sigma_{\;\,\mu}\right){\cal E}_{\nu\alpha}-\frac{13}{2}K_{\mu\beta}{\cal E}^\beta_{\;\,\alpha}K^{\alpha}_{\;\,\nu} \right.\nonumber\\\hspace{-1cm}&+&\left. \left(3\,K^\alpha_{\;\,\mu}K^{\beta}_{\;\,\alpha}-K_{\mu\alpha}K^{\alpha\beta}\right){\cal E}_{\nu\beta} - K_{\mu\alpha}K_{\nu\beta}{\cal E}^{\alpha\beta} - 4K^{\alpha\beta}R_{\mu\nu\gamma\alpha}K^{\gamma}_{\;\beta} - \frac{7}{6}K^{\sigma\beta}K^{\;\,\alpha}_{\mu}R_{\nu\sigma\alpha\beta}\Bigg]\,\frac{y^4}{4!} + \cdots \right.  \label{tay} \ea

Such expansion was analyzed in \cite{maartens} only up to the second order.  This procedure does not suffice to evince the additional terms, arising from the variable tension in the brane. For an alternative method which does not take into account the $\mathbb{Z}_2$ symmetry,  and some subsequent applications,  see \cite{Jennings:2004wz}.

There are still some additional terms that were concealed in the expression above, that are going to be emphatically focused on the next Sections, concerning the derivatives of the variable tension. The additional terms coming from the variable brane tension are shown to be essential for the subsequent analysis on the brane tension influence on the black string behavior along the extra dimension.

\subsection{Additional terms elicited from the variable tension}
Although the brane tension is variable, up to terms in $y^2$ at the expansion (\ref{tay}) there are no additional terms, which appear 
only from the order $|y|^3$ on, regarding (\ref{tay}). The unique term to contribute to the derivatives of the variable tension $\lambda$
comes from the coefficient ${\bf \pounds}_{\bf n}{\cal E}_{\mu\nu}$
of $|y|^3$ in (\ref{tay}), given by
$\nabla^\alpha
{\cal B}_{\alpha(\mu\nu)}$, wherein one can substitute the boundary condition $
{\cal B}_{\mu\nu\alpha} = 2\nabla_{[\mu}K_{\nu]\alpha}$ \cite{GCGR,maartens}. Apart the term related to the energy-momentum tensor in $K_{\mu\nu}=-\frac{1}{2}\kappa_5^2 \left[T_{\mu\nu}+ \frac{1}{3}
\left(\lambda-T\right)g_{\mu\nu} \right]$, since we are concerned only about the extra terms coming from the variable tension, such extra terms read:
\ba
\nabla^\alpha
{\cal B}_{\alpha(\mu\nu)}=-\frac{2}{3}\kappa_5^2\left((\nabla^\alpha\nabla_\alpha\lambda)g_{\mu\nu}-(\nabla_{(\nu}\nabla_{\mu)}\lambda)
\right).\label{addtruey3}
\ea

Terms in order $y^4$ can be obtained immediately, besides the additional terms coming from the variable tension brane, since we are interested on
the further effects of the variable brane tension on the black string character. Such extra terms in the order $y^4$ of expansion
(\ref{tay}) are given by
\ba
&&- \frac{1}{3}\kappa_5^2\,\left[\Box(\Box\lambda)g_{\mu\nu}-\nabla_{(\nu}\nabla_{\mu)}(\Box\lambda)\right]+\left(\frac{1}{3}\kappa_5^2+2 K\right)[(\Box\lambda){\cal E}_{(\mu\nu)} - \nabla^\alpha\left((\nabla_{(\mu}\lambda)\, {\cal E}_{\nu)\alpha}\right)]\nonumber\\
&&+6 \left[(\Box\lambda)K_{(\mu\sigma}{\cal E}_{\nu)}^\sigma - \nabla^\alpha((\nabla_{(\mu}\lambda)\, {\cal E}_{\nu)\alpha})\right]
+2\left(K + \frac{7}{3}\kappa_5^2\right)\left[(\Box\lambda)K\,K_{\mu\nu}-\nabla^\alpha((\nabla_{(\mu}\lambda)\, K\,K_{\alpha\nu)})\right]\nonumber\\
&&+ \frac{1}{3}\kappa_5^2 [(\Box\lambda)R_{\mu\nu}-\nabla^\alpha((\nabla_{(\mu}\lambda)\,R_{\alpha\nu)})]
-2 K^{\sigma\beta}\left[(\Box\lambda)R_{(\mu\sigma\nu)\beta} - \nabla^\alpha\left((\nabla_{(\mu}\lambda)\, R_{\alpha\sigma\nu)\beta}\right)\right]\nonumber\\&&+\left(2\,K_{\sigma\rho}K^{\sigma\rho}-\frac{1}{3}\Lambda_5 \right)[(\Box\lambda)g_{\mu\nu}-\nabla_{(\nu}\nabla_{\mu)}\lambda]\nonumber\\&&
+\frac{1}{3}\kappa_5^2\left[(\Box\lambda)(K_{(\mu\sigma}K_{\nu)\beta} K^{\sigma\beta} - (K_{\sigma\rho}K^{\sigma\rho})\,K_{(\mu\nu)}) - \nabla^\alpha\left((\nabla_{(\mu}\lambda)\, (K_{\alpha\sigma}K^{\;\,\sigma}_{\nu)} - K\,K_{\alpha\nu)})\right)\right]\,\label{addtruey4}
 \ea
and are obtained when one substitutes the Lie derivative of each term of such equation in the expression for $\lie\left({\bf \pounds}_{\bf n} {\cal E}_{\mu\nu}\right)$, taking into account once more Eqs.(\ref{exp1})-(\ref{exp4}).
We shall consider in this paper the brane tension as a time-dependent function $\lambda = \lambda(t)$. 

\subsection{Black string corrections for the vacuum case}

Now we focus on the situation where Equations (\ref{addtruey3}) and (\ref{addtruey4}) provide the further extra terms in the expansion given by Eq.(\ref{tay}) in the vacuum on the brane. In this case, $T_{\mu\nu}=0$, and consequently Eq.(\ref{ext}) is led to
 \be\label{ext1}
K_{\mu\nu}=-\frac{1}{6}\kappa_5^2\lambda g_{\mu\nu},
 \ee
 which provides the expansion at Eq.(\ref{tay}) to be written straightforwardly as
 \ba
g_{\mu\nu}(x,y)&=& g_{\mu\nu}-\frac{1}{3}\kappa_5^2\lambda g_{\mu\nu}\,|y|~~{}+\left[-{\cal E}_{\mu\nu} +\left(\frac{1}{36}\kappa_5^4\lambda^2 - \frac{1}{6}\Lambda_5\right)g_{\mu\nu}\right]\, y^2+ \nonumber\\
&& +\left(\left(-\frac{193}{216}\lambda^3\kappa_5^6 -\frac{5}{18}\Lambda_5\kappa_5^2\lambda\right)g_{\mu\nu} +\frac{1}{6}\kappa_5^2{\cal E}_{\mu\nu}+\frac{1}{3}\kappa_5^2({\cal E}_{\mu\nu}+R_{\mu\nu})\right)\,\frac{|y|^3}{3!} +
 \nonumber\\
&&+\left.\Bigg[\frac{1}{6}\Lambda_5\left(\left(R-\frac{1}{3}\Lambda_5 - \frac{1}{18}\lambda^2\kappa_5^4\right)+\frac{7}{324}\lambda^4\kappa_5^8\right)g_{\mu\nu} + \left(R -\Lambda_5 + \frac{19}{36}\lambda^2\kappa_5^4\right){\cal E}_{\mu\nu}\right.\nonumber\\
&&\left. \qquad + \left(\frac{37}{216}\lambda^2\kappa_5^4- \frac{1}{6}\Lambda_5\right)R_{\mu\nu}+ {\cal E}^{\alpha\beta}\,R_{\mu\alpha\nu\beta}\Bigg]\,\frac{y^4}{4!} + \cdots \right.\label{tay3}
\ea
 As the coefficients above concern uniquely quantities in the brane, we evince the property that when there is vacuum on the brane the
brane field equations
 \ba\label{vac}
R_{\mu\nu}=-{\cal E}_{\mu\nu},\qquad R^\mu {}_\mu =0={\cal E}^\mu
{}_\mu,\qquad \nabla^\nu {\cal E}_{\mu\nu}=0,\,
 \ea hold. It induces the last term of $|y|^3$ in Eq.(\ref{tay3}) equals zero.
Hence 
  Eq.(\ref{tay3}) is written as
 \ba
\hspace*{-.8cm}g_{\mu\nu}(x,y)&=&g_{\mu\nu}-\frac{1}{3}\kappa_5^2\lambda g_{\mu\nu}\,|y|~~{}+\left[-{\cal E}_{\mu\nu} +\left(\frac{1}{36}\kappa_5^4\lambda^2 - \frac{1}{6}\Lambda_5\right)g_{\mu\nu}\right]\, y^2\nonumber\\&&-\left(\left(\frac{193}{216}\lambda^3\kappa_5^6 +\frac{5}{18}\Lambda_5\kappa_5^2\lambda\right)g_{\mu\nu}+\frac{1}{6}\kappa_5^2{R}_{\mu\nu}\right)\,\frac{|y|^3}{3!} +
\nonumber\\&+&\left.\Bigg[\frac{1}{6}\Lambda_5\left(\left(R-\frac{1}{3}\Lambda_5 - \frac{1}{18}\lambda^2\kappa_5^4\right)+\frac{7}{324}\lambda^4\kappa_5^8\right)g_{\mu\nu}+ \left(R + \frac{5}{6}\Lambda_5 - \frac{77}{216}\lambda^2\kappa_5^4\right)R_{\mu\nu} - R^{\alpha\beta}\,R_{\mu\alpha\nu\beta}\Bigg]\,\frac{y^4}{4!} + \cdots \right.\label{tay31}
\ea

 Furthermore, in the vacuum Eq.(\ref{addtruey4}) --- corresponding to the additional terms arising from the derivatives of variable brane tension $\lambda$ --- is led to
\ba
&&- \frac{1}{3}\kappa_5^2\left(\Box(\Box\lambda)g_{\mu\nu}-\nabla_{(\nu}\nabla_{\mu)}\Box\lambda\right)+\left(-\frac{1}{3}\Lambda_5 +\frac{8}{9}\lambda^2\kappa_5^4\right)\left((\Box\lambda)g_{\mu\nu}-\nabla_{(\nu}\nabla_{\mu)}\lambda\right)\nonumber\\
&&+\frac{1}{9}\lambda^2\kappa_5^4(\lambda g_{\mu\nu}+\nabla_{(\nu}\nabla_{\mu)}\lambda)-\frac{4}{3}\kappa_5^2[(\Box\lambda){\cal E}_{(\mu\nu)} - \nabla^\alpha\left((\nabla_{(\mu}\lambda)\, {\cal E}_{\nu)\alpha}\right)]\label{vacexp}\\&&
+\frac{1}{3}\kappa_5^2(\Box\lambda)\left(\frac{5}{216}\lambda^3\kappa_5^6g_{\mu\nu}\right)
+\frac{1}{3}\lambda \kappa_5^2[(\Box\lambda)R_{(\mu\nu)} - \nabla^\alpha\left((\nabla_{(\mu}\lambda)\, R_{\alpha\nu)}\right)]\nonumber\\&&
+6 [(\Box\lambda)K_{(\mu\sigma}{\cal E}_{\nu)}^\sigma - \nabla^\alpha(\nabla_{(\mu}\lambda)\, {\cal E}_{\nu)\alpha}]\nonumber , \ea where on the brane $R_{\mu\nu} = -{\cal E}_{\mu\nu}$ holds as one of the field equations in (\ref{vac}).

\section{black string Schwarzschild corrections}

As the case of interest to be investigated is exactly the corrections on the black string horizon along the extra dimension, we shall focus on the term $g_{\theta\theta}$ --- corresponding to the square of the black string horizon --- of the expansion at Eq.(\ref{tay}). Clearly, a time dependent brane tension shall modify the black string Schwarzschild background. The complete solution is, however, hugely difficult to accomplish. Therefore we adopt an effective approach, studying the horizon variation with tension variation corrections only in the Taylor expansion. As shall be shown, even in this approximative case interesting results are accomplished.

A static spherical metric on the brane
can be expressed as \begin{equation}\label{124}
g_{\mu\nu}dx^{\mu}dx^{\nu} = - F(r)dt^2 + (H(r))^{-1}{dr^2} + r^2d\Omega^2,
\end{equation}
\noindent where $d\Omega^2$ denotes the  line-element of a 2-dimensional unit sphere.
The projected Weyl term on the brane is given by \cite{maartens}
\begin{eqnarray}
{\cal E}_{\theta\theta}&=&-1+H +\frac{r}{2}H\left(\frac{F'}{F}
+\frac{H'}{H} \right)=0,\label{weyl}
\end{eqnarray} for the Schwarzschild metric $F(r) = H(r)=\left(1-\frac{2M}{r}\right)$, where we denote $M \mapsto GM/c^2$.
As the black string horizon variation along the extra dimension is
analyzed, the term $g_{\theta\theta}(x,y)$ in (\ref{tay3}) above
is given by \ba
g_{\theta\theta}(x,y)&=&r^2-\frac{r^2}{3}\kappa_5^2\lambda\,|y|~~{}+\left(\frac{1}{36}\kappa_5^4\lambda^2 - \frac{1}{6}\Lambda_5\right)r^2\, y^2-\left(\frac{193}{216}\lambda^3\kappa_5^6 +\frac{5}{18}\Lambda_5\kappa_5^2\lambda\right)\,\frac{r^2|y|^3}{3!} \nonumber\\
&&   -\frac{1}{18}\Lambda_5\left(\left(\Lambda_5 + \frac{1}{6}\lambda^2\kappa_5^4\right)+\frac{7}{324}\lambda^4\kappa_5^8\right)\,\frac{r^2 y^4}{4!} +
\cdots\label{expp}
 \ea
Note that $g_{\theta\theta}=g_{\theta\theta}(x,y)$ coming from Eq. (14) is a component
of the Taylor expanded metric along the extra dimension, while
$g_{\mu\nu}$ in Eq. (16) depends only on the brane variables, as
usual \cite{maartens}. Now, as $\lambda = \lambda(t)$, the terms
$\nabla^\alpha\nabla_\alpha\lambda$ for the Schwarzschild metric
are computed and  the additional terms in  $|y|^3$ for
$g_{\theta\theta}$ in Eq.(\ref{expp}) are given by
  \ba
-\frac{2}{3}\kappa_5^2\lambda''r^2,\label{addtruey31}
\ea\noindent where $\lambda^\prime$ denotes the derivative with respect to the time $t$ --- and  the additional terms in
  $y^4$ for $g_{\theta\theta}$ are given by
\begin{eqnarray}&&
\left.\hspace*{-.59cm}\left(1-\frac{2M}{r}\right)\left[-\frac{1}{3}\kappa_5^2\left(1-\frac{2M}{r}\right)\lambda^{\prime\prime\prime\prime}r^2
+\lambda^{\prime\prime}r^2\left(\left(\frac{1}{3}\Lambda_5 -\frac{8}{9}\lambda^2\kappa_5^4\right)+ \frac{5}{648}\lambda^3\kappa_5^8\right)\right]
\right.\nonumber\\&+&\left. \frac{1}{9}\lambda^2\kappa_5^4\lambda r^2.\right. \label{expsch}  \ea All the expressions obtained are so far the most general. In order to better understand the physical implications of a variable tension in the event horizon along the extra dimension, let us particularize our analysis to a specific physical motivated case.

\subsection{A specific example regarding a brane variable tension}

In general, there are two distinct approaches to implement a variable tension brane. On the one hand, one may realize the brane tension as a (fundamental) scalar field appearing in the Lagrangian. This comprehensive picture is widely assumed in the context of string theory \cite{CORDAS} and supersymmetric branes \cite{SUSYBRANES}. Instead, to emulate such approach  the brane tension may be understood as an intrinsic property of the brane as in, e. g., \cite{GERGELY2008,GERGELY2009,PRD,PRDII,EPJC}. We delve into this point of view in our subsequent analysis.

From the braneworld picture, the functional form of the variable tension on the brane is an open issue. However, taking into account the huge variation of the universe temperature during its cosmological evolution, it is indeed plausible to implement the brane tension as a variable function of spacetime coordinates. In particular, in such case as a function of time coordinate. Although it lacks a complete scenario still,  the phenomenological interesting case regarding E\"otv\"os standard fluid membranes \cite{Eotvos} is useful to extract deep physical results.

The phenomenological E\"otv\"os law asserts that the fluid membrane depends on the temperature as \be \lambda=\chi (T_{c}-T),\label{TERM0}\ee where $\chi$ is a constant and $T_{c}$ represents a critical temperature denoting the highest temperature for which the membrane exists. We  heuristically depict in what follows how the brane tension varies, in full compliance with E\"otv\"os membranes models.

If there are no stresses in the bulk --- apart  the cosmological constant --- there is no exchange of energy-momentum between the bulk and the brane \cite{maartens}. Herewith it is possible to assert that there is no exchange of heat between the brane and the bulk --- the regarded interaction is purely gravitational. Therefore the well known expression $dQ=dE+pdV=0$ holds for the brane, and concerning photons from the cosmological microwave background the explicit formul\ae\,  $E=E_{\gamma}=\sigma T\el{4}V$ and $p=E/3=\frac{\sigma T\el{4}}{3}V$ accrue. Using these relations it is straightforwardly  verified that \be \frac{1}{T}\frac{dT}{dt}=-\frac{1}{3V}\frac{dV}{dT}. \label{TERM2} \ee Now, relating the volume to the Friedmann-Robertson-Walker scale factor  --- $V=a\el{3}(t)$ --- it follows that $T \sim \frac{1}{a(t)}$, and therefore one may use a physically motived input regarding the tension variable functions.

A complete cosmological setup must be forthwith obtained from the solution of the full cosmological brane equations taking into account variable tension, which is in general obviously a hard task. Notwithstanding we are assuming that the variability of tension may be outlined by considering the relation $T\sim a\el{-1}(t)$ valid in some approximation \cite{GERGELY2009}. This conservative approach, although not the most complete one, is certainly in full compliance with the standard cosmological model. Therefore, considering the analogy to E\"otv\"os membranes and the expression $T\sim a\el{-1}(t)$ we have $\lambda=\lambda_{0}(1-a_{min}/a)$, where $\lambda_{0}$ is a constant related to the $4$-dimensional coupling constants \cite{GERGELY2009}, and $a_{min}$ denotes the minimum scale factor under which the brane cannot exist, since its tension would become negative. For the qualitative analysis that we develop it is sufficient to consider the general shape of $\lambda$ given by
\be \lambda=1-\frac{1}{a(t)},\label{MOD2} \ee with normalized tension and scale factor. In order to supplement our assumption in (\ref{MOD2}), we stress that this type of tension variation may be useful to reconcile supersymmetry and inflationary cosmology. Indeed, as first noticed in Ref. \cite{GERGELY2009}, the time dependent brane tension gives rise to a time dependent $4$-dimensional effective cosmological constant. In particular for a tension varying as Eq.(\ref{MOD2}) it is possible to show that \be \Lambda_{4D}\sim 1-\frac{1}{a(t)}\left(1-\frac{1}{a(t)}\right).\label{MOD3}\ee  Hence, the cosmological constant starts at high negative values and as the brane universe expands it converges to a positive small value. It is indeed a remarkable characteristic of this type of tension\footnote{For another model presenting such a behavior of the effective cosmological constant see \cite{BARVINSKY}.}.

To finalize, a de Sitter-like brane behavior is assumed, by setting $a\sim e\el{\alpha t}$ with positive $\alpha$ \cite{RE41}, in such way that \be \lambda(t)=1-e\el{-\alpha t}.\label{ULTI}\ee It is important, in view of the assumption (\ref{ULTI}), to assert few remarks concerning its phenomenological viability. It is well known that the projected gravitational constant depends linearly on the brane tension \cite{maartens,GCGR}. Therefore, a time variable tension engenders a variation on the Newtonian constant $G\sim \lambda(t)$. The variation of a dimensional constant may always be incorporated in a suitable redefinition of length, time and energy \cite{REDE}. Nevertheless, it is possible to pick up some arbitrary value defining it as the standard one, and study its possible fractional variation. The recent astrophysical data indicate the constancy of the Newtonian constant (for an up to date review see \cite{WILL}). In fact, the best model independent bound on $\dot{G}/G$ is given by lunar laser ranging measurements whose upper limit is $(4\pm 9)\times 10\el{-13} yr\el{-1}$. Hence, taking into account Eq.(\ref{ULTI}), the following condition \be e\el{\alpha t}>1+10\el{13}\alpha\label{SN1} \ee must hold. In this qualitative analysis we shall not fix the value of $\alpha$, but as it is not desirable a huge departure from the standard scenario, it is expected to have $\alpha \sim H_{0}$, the Hubble parameter \cite{maartens}. Thus, in principle, $10\el{13}\alpha$ is not so far from $1$. An important characteristic of the constraint (\ref{SN1}) is that there is a particular time, say $\bar{t}$, given by $\bar{t}=\ln\big(1+10\el{13}\alpha\big)/\alpha$ below which this constraint is violated. This apparent difficulty may be circumvented by the fact that it is possible to introduce free parameters on the functional form of $\lambda$ such that $\bar{t}$ is small enough, making the condition $t<\bar{t}$ to belong to an early time range, when the assumption $T\sim 1/a(t)$ -- possibly --  no longer holds anymore. Besides, it is possible to interpret the violation of $(\ref{SN1})$ as a change variation faster than $10\el{-13}\,yr^{-1}$, which could in principle be scrutinized at intermediate redshifts \cite{NO}.

Having explicitly presented the type of variable tension, let us particularize our analysis taking into account Eq.(\ref{ULTI}). For the E\"otv\"os brane scenario, the following figures show the black string behavior along the extra dimension $y$ in the Schwarzschild picture, as predicted by Eqs.(\ref{expp}, \ref{addtruey31}, \ref{expsch}). The graphics below for each figure illustrate how the black string horizon varies along the extra dimension. Eq.(\ref{expp}) is the landmark for all graphics hereon, which we imposed 
$r=1 = \Lambda_5 = \kappa_5$, in order to make our analysis straightforward. It is clear that such re-scaling does not affect the shape of the graphics below.
\begin{figure}[H]
\begin{center}\includegraphics[width=3.85in]{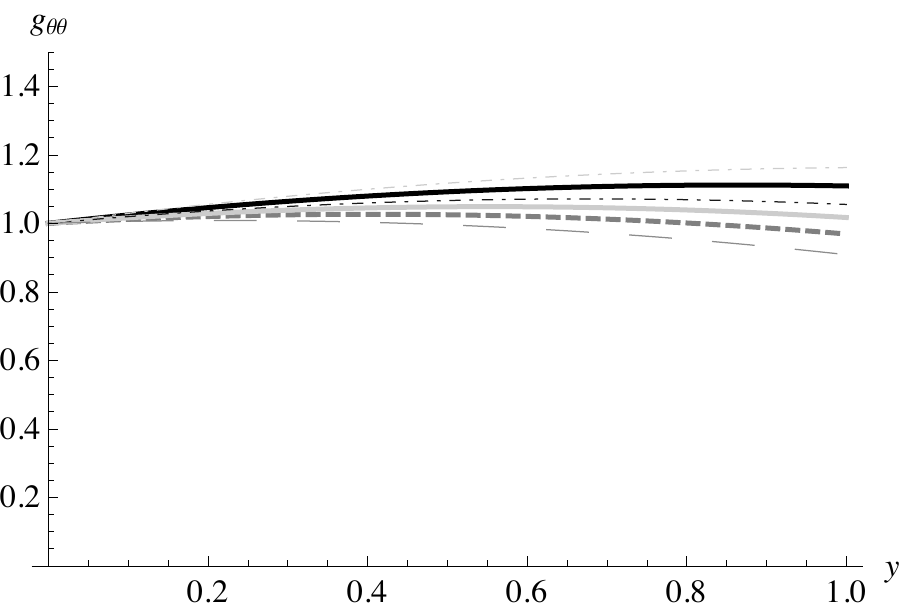}
\caption{\small Graphic of the brane effect-corrected  black string horizon $g_{\theta\theta}$ for the Schwarzschild metric with variable tension, along the extra dimension $y$ and also as function of the time $t$. The brane tension is given by $\lambda(t) = 1 - \exp(-\alpha t)$. We included, for comparison criteria, merely terms up to there order $y^2$ in the expansion (\ref{expp}). For the long dashed gray line (lower curve):  $\alpha t = 0.25$; for the dashed thick dark gray line: $\alpha t = 0.5$; for the  gray thick line $\alpha t = 0.75$; for the dot dashed line:  $\alpha t = 1$; for the  black line:  $\alpha t = 1.5$; for the dot dashed light gray line:  $\alpha t = 2.5$.  }
\end{center}
\end{figure}
\begin{figure}[H]
\begin{center}\includegraphics[width=3.5in]{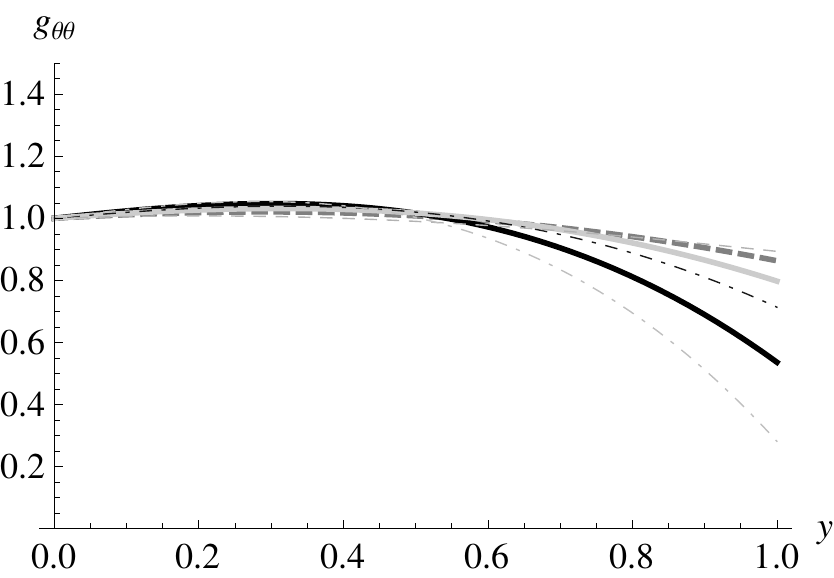}
\caption{\small Graphic of the brane effect-corrected  black string horizon $g_{\theta\theta}$ for the Schwarzschild metric with variable tension, along the extra dimension $y$ and also as function of the time $t$. The brane tension is given by $\lambda(t) = 1 - \exp(-\alpha t)$. This graphic \emph{does not} take into account the extra terms given by Eqs.(\ref{addtruey31}) and (\ref{expsch}). For the dashed light gray line:  $\alpha t = 0.25$; for the dashed thick dark gray line: $\alpha t = 0.5$; for the  gray thick line $\alpha t = 0.75$; for the dot dashed line:  $\alpha t = 1$; for the  black line:  $\alpha t = 1.5$; for the dot dashed light gray line:  $\alpha t = 2.5$.}
\end{center}
\end{figure}
\begin{figure}[H]
\begin{center}\includegraphics[width=3.65in]{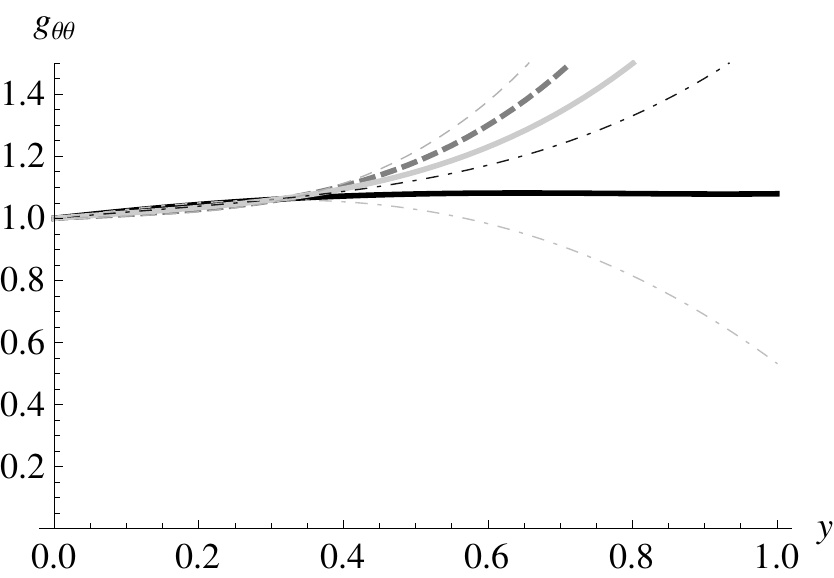}
\caption{\small Graphic of the brane effect-corrected  black string horizon $g_{\theta\theta}$ for the Schwarzschild metric with variable tension, along the extra dimension $y$ and also as function of the time $t$. The brane tension is given by $\lambda(t) = 1 - \exp(-\alpha t)$. This graphic \emph{does} take into account the extra terms given by Eqs.(\ref{addtruey31}) and (\ref{expsch}). For the dashed light gray line:  $\alpha t = 0.25$; for the dashed thick dark gray line: $\alpha t = 0.5$; for the  gray thick line $\alpha t = 0.75$; for the dot dashed line:  $\alpha t = 1$; for the  black line:  $\alpha t = 1.5$; for the dot dashed light gray line:  $\alpha t = 2.5$.}
\end{center}
\end{figure}

In Figures 1-3  the black string horizon behavior is obviously
different for  differing values of $\alpha t$. In Figure 1, for the sake of completeness we depict the
black string horizon behavior along the extra dimension, but now
considering only terms up to the order $y^2$ in Eq.(\ref{tay}), commonly approached
in the literature, for instance, in \cite{maartens}. 
 Figures 2 and 3 show  the variable tension brane correction including all terms up to order $y^4$, respectively without and with the terms elicited in Eqs.(\ref{addtruey31}) and (\ref{expsch}), regarding the derivatives of the brane variable tension. Those
graphics show the paramount importance of considering more terms
in the metric expansion given by Eq.(8), as accomplished
heretofore.   It also robustly illustrates that those terms drastically modify the black string horizon along the extra dimension, in the model here presented. 

\begin{figure}[H]
\begin{center}\includegraphics[width=2.9in]{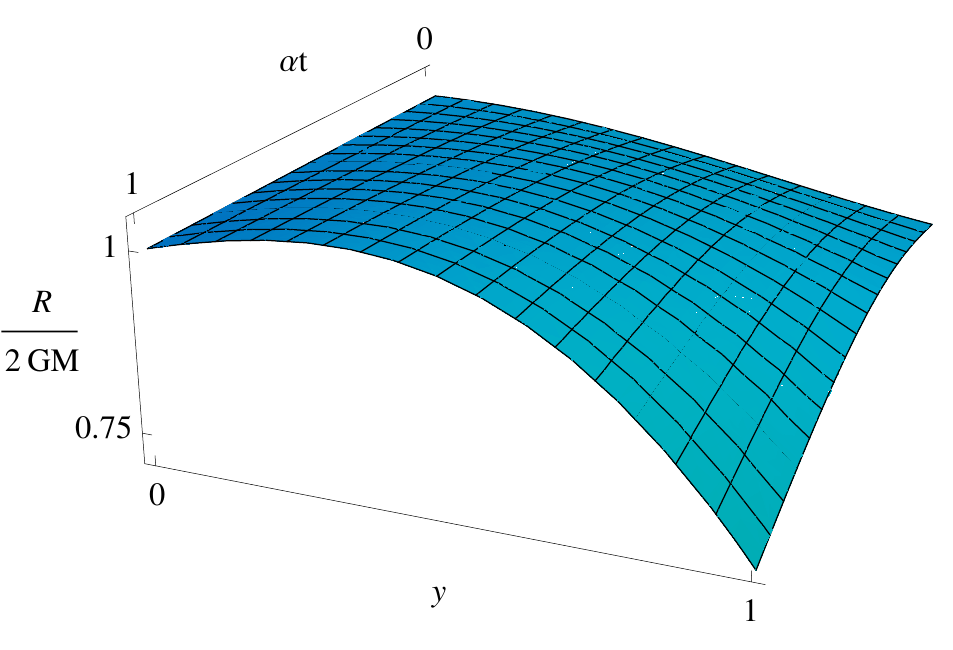}
\caption{\small Graphic of the brane effect-corrected  black string horizon $g_{\theta\theta}$ for the Schwarzschild metric with variable tension, along the extra dimension $y$ and also as function of the time $t$. The brane tension is given by $\lambda(t) = 1 - \exp(-\alpha t)$. This graphic \emph{does not} take into account the extra terms given by Eq.(\ref{addtruey4}, \ref{addtruey31}), that for the case considered are encrypted in Eq.(\ref{expsch}). }
\end{center}
\end{figure}

\begin{figure}[H]
\begin{center}\includegraphics[width=2.9in]{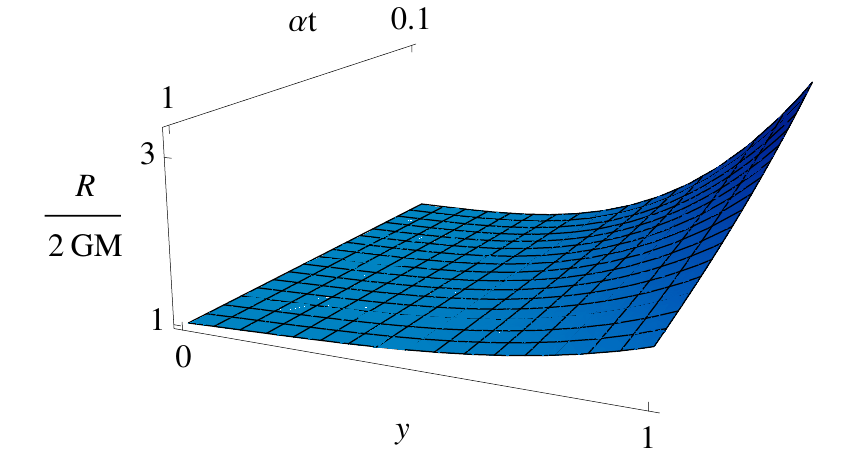}
\caption{\small Graphic of the brane effect-corrected  black string horizon $g_{\theta\theta}$ for the Schwarzschild metric with variable tension, along the extra dimension $y$ and also as function of the time $t$. The brane tension is given by $\lambda(t) = 1 - \exp(-\alpha t)$. This graphic \emph{does} take into account the extra terms given by Eq.(\ref{addtruey4}, \ref{addtruey31}), that for the case considered are encrypted in Eq.(\ref{expsch}). The black string normalized horizon varies from $R/(2GM)$ to $R=3/(2GM)$ approximately.}
\end{center}
\end{figure}

For the 3D graphics, Figure 4 does not take into account the time
derivative terms, while Figure 5 does take. We shall draw some
remarks in details on the general behavior encoded in the figures in the next
Section.

\section{Concluding Remarks}

In this paper branes whose variable tension is only dependent on
the time are focused. It is provided the Taylor expansion of the
metric along the extra dimension for this case.

Albeit terms up to the second order already evince the variable
tension character, when terms from the third order on are
investigated, it is throughly possible to unveil the real
character of the variable tension concerning the changing of the
event horizon. It accrues additional and unexpected properties,
since only in this case the covariant derivatives of the variable
tension appear, in the metric expansion along the extra dimension.
Furthermore, it is prominently necessary to consider such terms,
since in some regions of spacetime described by a brane the value
of the tension can be very small, and at the same time the brane
tension can bluntly vary along the spacetime coordinates.
Additional terms in the expansion analyzed modify the rate of the
horizon variation along the extra dimension, when the brane
tension varies in a braneworld model based upon the E\"otv\"os
law.

This unexpected feature on this effective model is explicitly
shown in Figures 1-5. In the first Figure the brane
effect-corrected  black string horizon $g_{\theta\theta}$ for the
Schwarzschild metric with variable tension is depicted along the
extra dimension $y$. In Figure 1 the
black string horizon behavior along the extra dimension is illustrated, regarding only terms up to the order $y^2$, commonly approached
in the literature, for instance, in \cite{maartens}. Figure 2 shows that the black string behavior
is obviously different for different fixed moments without taking
into account terms of time derivative of the tension, while Figure
3 does take into account such terms.

By comparing Figures 4 and 5, it is possible to see that in
the Figure 5, for all $t \geq 0$, the black string presents no
singularities. It is precluded by the extra terms in the Taylor
expansion of the metric in the brane-corrected black string
horizon in our model. Those terms do not appear in models with
constant brane tension, since they encode the covariant
derivatives of the brane variable tension $\lambda$.

The general result encoded in the figures is exhaustive: whenever
the time derivative terms are taken into account the variation of
the horizon along the extra dimension is such that the horizon
does not tend to zero. This effect is, presumable, naively
interpreted within this (eminently) classical framework. Still,
one could speculate that it would be interpreted in terms of
fluctuations around the brane. In fact, combing the fact that a
completely rigid object cannot exist in the general relativity
framework with the presence of a scalar field representing the
brane position into the bulk, one arrives at the possibility of a
spontaneous symmetry breaking of the bulk diffeomorphism. In this
way, a perturbative spectrum of scalar particles, the so called
branons, may appear if the tension scale is much smaller then the
higher dimensional mass scale \cite{BRANON}.

Now, being the brane tension a variable quantity, it is expected a
non-trivial contribution to the branons production. Besides, the
tension derivatives (computing the rate of tension variation) may
also have an important role in the branons production. Following
this reasoning, one could guess that such fluctuations could (in
principle) to supply the black hole horizon along the extra
dimension horizon, what makes its approach to the singularity
difficult. Obviously, this interpretation must be enforced by a
critical analysis of the precise influence of a variable tension
on the branons production, as well as the branons influence on the
black string behavior. These issues together with possible quantum
effects \cite{ruth33} are currently under investigation.

\section*{Acknowledgments}
R. da Rocha is grateful to Conselho Nacional de Desenvolvimento Cient\'{\i}fico e Tecnol\'ogico (CNPq)  476580/2010-2 and
304862/2009-6 for financial support.


\begin{thebibliography}{99}
\footnotesize

\bibitem{HW} P. Horava, E. Witten, \emph{Heterotic and type I string dynamics from eleven-dimensions, Nucl. Phys. B} {\bf 460} (1996) 506;  P. Horava and E. Witten, \emph{Eleven-dimensional supergravity on a manifold with boundary, Nucl. Phys. B} {\bf 475} (1996) 94.

\bibitem{PREV} V. A. Rubakov, M. E. Shaposhnikov, \emph{Do we live inside a domain wall?, Phys. Lett. B} {\bf 125} (1983) 136; V. A. Rubakov, M. E. Shaposhnikov, \emph{Extra space-time dimensions: towards a solution to the cosmological constant problem, Phys. Lett. B} {\bf 125} (1983) 139; K. Akama, \emph{Pregeometry, Lecture Notes in Physics, vol. 176, Gauge Theory and Gravitation, Proceedings, Nara, 1982, ed. by Kikkawa, N. Nakanishi, H. Nariai} (Springer, Heidelberg, 1983) pp. 267-271.

\bibitem{nossos}  J. M. Hoff da Silva, R. da Rocha, \emph{Braneworld Remarks in Riemann-Cartan Manifolds, Class. Quant. Grav.} {\bf 26} (2009) 055007 [{\tt arXiv:0804.4261v4 [gr-qc]}] [Erratum ibidem {\bf 26} (2009) 179801].

\bibitem{adsbranes} J. M. Hoff da Silva, R. da Rocha, \emph{Gravitational constraints of dS branes in AdS Einstein-Brans-Dicke bulk, Class. Quant. Grav.} {\bf 27} (2010) 225008 [{\tt arXiv:1006.5176v1 [gr-qc]}].

\bibitem{prd2010} J. M. Hoff da Silva, R. da Rocha, \emph{Torsion Effects in Braneworld Scenarios, Phys. Rev. D} {\bf 81} (2010) 024021 [{\tt arXiv:0912.5186v1 [hep-th]}].

\bibitem{ROLDAO/CARLAO} R. da Rocha, C. H. Coimbra-Araujo, \emph{Extra dimensions in CERN LHC via mini-black holes: effective Kerr-Newman brane-world effects, Phys. Rev. D} {\bf 74} (2006) 055006
[{\tt arXiv:hep-ph/0607027v3}];  R. da Rocha, C. H. Coimbra-Araujo, \emph{Variation in the luminosity of Kerr quasars due to extra dimension in brane. Randall-Sundrum model, JCAP} {\bf 12} (2005) 009 [{\tt arXiv:astro-ph/0510318v2}]; C. H. Coimbra-Araujo, R. da Rocha, I. T. Pedron, \emph{Anti-de Sitter curvature radius constrained by quasars in brane-world scenarios,   Int. J. Mod. Phys. D} {\bf 14} (2005) 1883 [{\tt arXiv:astro-ph/0505132v4}]; R. da Rocha, C. H. Coimbra-Araujo, \emph{Physical Effects of Extra Dimension and Concomitant Map between Photons and
Gravitons in Randall-Sundrum Brane-World Scenario, PoS} {\bf IC2006} (2006) 065 [{\tt arXiv:gr-qc/0610134v1}]; R.~da Rocha, J.~M.~Hoff da Silva, \emph{Torsion influence in braneworld scenarios,
  PoS} {\bf ISFTG} (2009) 026.

\bibitem{nova} J. M. Hoff da Silva, Roldao da Rocha,
\emph{Possible Generalizations within Braneworld Scenarios: Torsion fields}, Chapter of ``Classical and Quantum Gravity: Theory, Analysis and Applications", Vincent R. Frignanni (Editor), Nova Science Pub. Inc., Hauppauge (2011) [{\tt arXiv:1012.2108v1 [gr-qc]}].


\bibitem{Gergely:2006hd}
  L.~A.~Gergely,
  \emph{Black holes and dark energy from gravitational collapse on the brane,}
  \emph{JCAP} {\bf 0702 } (2007)  027
  [{\tt  arXiv:hep-th/0603254}].


\bibitem{loo}
  M.~Kavic, J.~H.~Simonetti, S.~E.~Cutchin, S.~W.~Ellingson, C.~D.~Patterson,
  \emph{Transient Pulses from Exploding Primordial Black Holes as a Signature of an Extra Dimension,
  JCAP} {\bf 0811 } (2008)  017,
  [{\tt arXiv:0801.4023 [astro-ph]}].

\bibitem{Anderson:2005af}
  E.~Anderson, R.~Tavakol,
  \emph{Geodesics, the equivalence principle and singularities in higher-dimensional general relativity and braneworlds},
  \emph{JCAP} {\bf 0510 } (2005)  017.
  [{\tt arXiv:gr-qc/0509055 [gr-qc]}].

\bibitem{casadio1} R. Casadio, C. Germani, \emph{Gravitational collapse and black hole evolution: do holographic black holes eventually "anti-evaporate"?}, Prog. Theor. Phys. {\bf 114} (2005) 23-56 [{\tt
arXiv:hep-th/0407191v4}].

\bibitem{VACARU} S. I. Vacaru, in {\it Clifford and Riemann Finsler Structures in Geometric Mechanics and Gravity}, selected works by S. Vacaru, edited by P. Stavrinos, E. Gaburov, and D. Gonta, Geometry Balkan Press, Bucharest, 2006, Chap. 7.

\bibitem{GERGELY2008} L. A. Gergely, \emph{Friedmann branes with variable tension, Phys. Rev. D} {\bf  78} (2008) 084006 [{\tt arXiv: 0806.3857v3 [gr-qc]}].

\bibitem{GERGELY2009} L. A. Gergely, \emph{E\"otv\"os branes, Phys. Rev. D} {\bf  79} (2009) 086007 [{\tt arXiv: 0806.4006v2 [gr-qc]}].

\bibitem{PRD} M. C. B. Abdalla, J. M. Hoff da Silva, R. da Rocha, \emph{Notes on the Two-brane Model with Variable Tension, Phys. Rev. D} {\bf 80} (2009) 046003 [{\tt arXiv: 1101.4214v1 [gr-qc]}].

\bibitem{PRDII} J. M. Hoff da Silva, \emph{Two-branes with variable tension model and the effective Newtonian constant, Phys. Rev. D} {\bf 83} (2011) 066001 [{\tt arXiv:0907.1321v1 [hep-th]}].

\bibitem{JMHEP} M. C. Abdalla, M. E. X. Guimar\~aes, J. M. Hoff da Silva, \emph{Positive tension 3-branes in an  AdS$_ {5}$ bulk, JHEP} {\bf 09} (2010) 051 [{\tt arXiv:1001.1075v2 [hep-th]}].

\bibitem{EPJC}  K. C. Wong, K. S. Cheng, T. Harko, \emph{Inflation and late time acceleration in braneworld cosmological models with varying brane tension, Eur. Phys. J. C} {\bf 68} (2010) 241 {[\tt arXiv:1005.3101v1 [gr-qc]}].

\bibitem{maartens} R. Maartens, \emph{Brane-world gravity, Living Rev. Relat.} {\bf 7}, 7 (2004) [{\tt arXiv:gr-qc/0312059v2}].

\bibitem{RSI} L. Randall and R. Sundrum, \emph{A Large Mass Hierarchy from a Small Extra Dimension, Phys. Rev. Lett.} {\bf 83} (1999) 3370 [{\tt arXiv:hep-ph/9905221v1}].


\bibitem{1} G. F. R. Ellis, \emph{Carg\'ese Lectures in Physics vol. VI} E. Schatzman (ed.) Gordon and Breach, New York 1973.

\bibitem{GCGR} T. Shiromizu, K. Maeda, and M. Sasaki, \emph{The Einstein Equations on the 3-Brane World, Phys. Rev. D} {\bf 62} (2000) 043523 [{\tt arXiv:gr-qc/9910076v3}]; A. N. Aliev and A. E. Gumrukcuoglu, \emph{Gravitational Field Equations on and off a 3-Brane World, Class. Quant. Grav.} {\bf 21} (2004) 5081 [{\tt arXiv:hep-th/0407095v1}].

\bibitem{Gergely:2003pn}
  L.~A.~Gergely, \emph{Generalized Friedmann branes,
  Phys.\ Rev.\  D} {\bf 68} (2003) 124011
  [{\tt arXiv:gr-qc/0308072}].

 \bibitem{Jennings:2004wz}
  D.~Jennings, I.~R.~Vernon, A.~-C.~Davis, C.~van de Bruck,
  \emph{Bulk black holes radiating in non-Z(2) brane-world spacetimes,
  JCAP} {\bf 0504 } (2005)  013.
  [{\tt arXiv:hep-th/0412281}].

\bibitem{CORDAS} E. Bergshoeff and P. K. Towsend, \emph{Super D-branes revisited, Nucl. Phys. B} {\bf 531} (1998) 226 [{\tt arXiv:hep-th/9804011v2}].

\bibitem{SUSYBRANES} E. Bergshoeff, R. Kallosh, and A. Van Proeyen, \emph{Supersymmetry in Singular Spaces, JHEP} {\bf 10} (2000) 033 [{\tt arXiv:hep-th/0007044v3}].

\bibitem{Eotvos} R. E\"otv\"os, Wied. Ann. {\bf 27} (1886) 448.

\bibitem{BARVINSKY} A. O. Barvinsky, C. Deffayet, and A. Yu. Kamenshchik, \emph{CFT driven cosmology and the DGP/CFT correspondence, JCAP} {\bf 05} (2010) 034 [{\tt arXiv:0912.4604v1 [hep-th]}].

\bibitem{REDE} C. J. A. P. Martins, E. Menegoni, S. Galli, G. Mangano, and A. Melchiorri, \emph{Varying couplings in the early universe: correlated variations of $\alpha$ and G, Phys. Rev. D} {\bf 82} (2010) 023532 [{\tt arXiv:1001.3418}].

\bibitem{WILL} C. M. Will, \emph{The confrontation between General Relativity and experiment, Living Rev. Relativity} {\bf 9} (2005) 3 [{\tt arXiv:gr-qc/0510072v2}].

\bibitem{NO} N. Yunes, F. Pretorius, and D. Spergel, \emph{Constraining the evolutionary history of Newton's constant with gravitational wave observations, Phys. Rev. D} {\bf 81} (2010) 064018 [{\tt arXiv:0912.2724v2 [gr-qc]}].

\bibitem{RE41} A. Campos, C. F. Sopuerta, \emph{Evolution of Cosmological Models in the Brane-world Scenario, Phys. Rev. D} {\bf 63} (2001) 104012 [{\tt arXiv:hep-th/0101060v1}]; A. Campos, C. F. Sopuerta, \emph{Bulk effects in the cosmological dynamics of brane-world scenarios, Phys. Rev. D} {\bf 64} (2001) 104011 [{\tt arXiv:hep-th/0105100v1}].

\bibitem{BRANON} M. Bando, T. Kugo, T. Noguchi, and K. Yoshioka, \emph{Brane fluctuations and suppression of Kaluza-Klein mode couplings, Phys. Rev. Lett.} {\bf 83} (1999) 3601 [{\tt arXiv:hep-ph/9906549v2}]; J. A. R. Cembranos, A. Dobado, and A. L. Maroto, \emph{Brane-World dark matter, Phys. Rev. Lett.} {\bf 90} (2003) 241301 [{\tt arXiv:hep-ph/0302041}].


\bibitem{ruth33} R. Gregory, R. Whisker, K. Beckwith, and C. Done, \emph{Observing braneworld black holes},
\emph{JCAP} {\bf 1004} (2004) 013 [{\tt arXiv:hep-th/0406252v1}].




\end{thebibliography}
\end{document}